\documentclass[12pt]{article}
\usepackage{latexsym}
\usepackage{graphicx}
\usepackage[usenames,dvipsnames]{color}

\def\be{\begin{equation}}
\def\ee{\end{equation}} 
\def\bea{\begin{eqnarray}}
\def\eea{\end{eqnarray}}

\def\e{\emph}

\def\nb{$N$-body problem }
\def\nbn{$N$-body problem}

\def\mb{$N$-body }

\def\case#1/#2{\textstyle\frac{#1}{#2}}

\def\e{\emph}

\def\s{${\mathcal S}$ }
\def\sn{${\mathcal S}$}

\def\tb{three-body problem }

\def\el{$\ell_\textrm{\scriptsize rms}$ }
\def\eln{$\ell_\textrm{\scriptsize rms}$}

\def\npn{$V_\textrm{\scriptsize shape}$}
\def\c{$C_\textrm{\scriptsize shape}$ }
\def\cn{$C_\textrm{\scriptsize shape}$}
\def\isa{$V_\textrm{\scriptsize New}$ }
\def\isan{$V_\textrm{\scriptsize New}$}
\def\po{Poincar\'e }

\def\n{Newtonian }

\begin{document}

\begin{center}

{\bf \Huge{Gravity's Creative Core}}

\vspace{.3in}

{\bf \Large{Julian Barbour}}\footnote{julian.barbour@physics.ox.ac.uk}

\vspace{.1in}

College Farm, The Town, South Newington, Banbury, OX15 4JG, UK.

\end{center}

\abstract{I argue that the essence of gravity can be understood only in the context of the universe and that unrecognised implicit retention of Newtonian absolute scale and the impact of thermodynamics have obscured it. Typical attempts to resolve the conflict between maximal matter entropy in the early universe and the second law of thermodynamics illustrate my case. Rovelli, for example, argues that the scale factor, and with it the overall state, was out of equilibrium. He illustrates subsequent entropy-increasing interaction with other degrees of freedom in two models: particles in a box interacting with a piston initially out of equilibrium and Newtonian gravitating particles. However, the piston's position and momentum, like the particles', are defined relative to the box, while unobservable absolute space defines those of the gravitating particles. Their representation using observable scale-invariant variables shows that the box and gravitational statistics differ greatly: in the latter a single degree of freedom, gravity's creative core, drives the system in every solution from entropic-like disorder to ever increasing ordered structure. Since in many ways such particles are a good approximation to general relativity, this may also be true for our universe.}

\section{Introduction\label{a}}

As noted in \cite{grub}, it is striking that virtually all discussions of the second law of thermodynamics considered in the context of the universe employ, largely unchanged, mathematical frameworks and concepts that arose from study of systems within the universe, not of the universe itself. Thus the study of irreversibility is seldom extended beyond the framework created by Carnot to study the efficiency of steam engines, which stop if the steam escapes from their cylinders. Thermodynamics is the science of confined systems and equilibration within them.\footnote{The statistics of systems that are closed in the sense of not being subject to external forces can differ significantly from those of ones that are confined, either physically by a box in a laboratory or theoretically by an artificial nondynamical potential wall. The \nb discussed in this paper is closed in the traditional sense but not confined because its scale variable can grow without bound. General relativity has spatially closed solutions but, if eternally expanding, their volume and Hubble radius grow without bound.} To provide an atomistic explanation of phenomenological thermodynamics, the founders of statistical mechanics accordingly always considered particles in a box, the walls of which prevented their escape. Such systems have phase spaces of bounded Liouville measure. Despite reading much that relates to the second law, irreversibility, and the arrow of time, I have nowhere seen serious discussion of whether statistical effects are likely to be different in systems with phase spaces of bounded or unbounded measure.\footnote{However, in his classical study of the statistical-mechanical foundation of thermodynamics \cite{gibbs} Gibbs comments that in treating of the canonical distribution he will always assume the corresponding integral ``has a finite value, as otherwise \dots~the law of distribution becomes illusory.'' This has the consequence that his method cannot treat instances ``in which the system or parts of it can be distributed in unlimited space'' and matches the fact that in thermodynamics there is ``no thermodynamic equilibrium of a (finite) mass of gas in an infinite space''  \cite{gibbs}, p.~35. He also notes that there are necessary restrictions on the momenta. Gibbs thus indicates that his probabilistic statistical-mechanical derivation of the laws of thermodynamics can only apply to systems with a bounded phase space. He does not discuss how and to what extent statistical arguments might apply to systems with unbounded phase spaces.}  It is true that in ``Where was low past entropy?''~\cite{rov} on which I comment here Rovelli does note that in conventional thermodynamics a considered system can be acted on by external agents and that this is not so for the universe. However, neither in \cite{rov} nor as the physicist in \cite{brov} does Rovelli discuss whether the universe has a phase space of bounded or unbounded measure and what the implications of the difference might be. Moreover, in \cite{rov} he equates conceptually the statistics of particles in a box interacting with a spring-powered piston and the statistics of the \n \nbn. While I agree with Rovelli and Wallace \cite{wall}, whom Rovelli cites, that the \nb is, as regards time's arrow, likely to be a good guide to the behaviour of the universe, I question the parallel with the piston and particles in a box. 

Closely related to this is implicit retention of absolute scale that came into physics with Newton's concept of absolute space. This affects the discussion of not only \cite{rov, wall} but also other authors, including Penrose \cite{pene, penr}. Thus, both Penrose \cite{penr}, p.~701, and Wallace \cite{wall}, p.~20, citing Penrose, say explicitly that in cosmology small-scale local degrees of freedom (dofs) interact with the universe's large-scale size dof. This presupposes an external ruler, which cannot be right. Only ratios of  observable quantities have meaning in physics. A universe of gravitating particles must be described solely in terms of ratios of the separations between them. In the simplest case of three particles, only the ratios of the side lengths of the triangle they form are meaningful. They determine the internal angles and with them the triangle's shape.

Here, Heisenberg's single-sentence abstract of his 1925 paper that created matrix mechanics may be cited. It highlights the transformative power of treatment in terms of observables. In translation it reads: ``The present paper seeks to establish a basis for theoretical quantum mechanics founded exclusively upon relationships between quantities which in principle are observable.'' Heisenberg's aspiration, achieved with spectacular success, reveals a clear influence of Mach's epistemology, though Heisenberg applied it in practice only to quantities observed in the laboratory whereas Mach applied it to the universe, of which he repeatedly said it should be described exclusively in terms of relative separations \cite{mach}.\footnote{Einstein famously criticized Heisenberg for thinking a theory could be based exclusively on observable quantities \cite{h}. However, as Heisenberg reports, Einstein's main concern was with Mach's general principle of science and he granted that he himself had used such a philosophy and that it could have heuristic value. When looking at the stars in the sky on a dark night, we all have the direct experience of seeing them separated by angles and from experiences like that, above locating ourselves relative to visible objects and not unobservable space, it is natural to suppose that separations between bodies, not positions in unobservable space, are fundamental. Such was Mach's view and before him many others, above all Leibniz. The interpretation of what is seen in an electron microscope may indeed be `theory-laden' but the experience of what is seen is not and is surely a valid stimulus to theory creation.}

There being no ruler outside the universe, we need to go beyond Mach and base kinematics, not on separations, but on \e{ratios of separations}, which define the shape of a finite set of particles. Note also that the origin of inertial frames of reference, Mach's concern, and the origin  of time's arrow, Rovelli's concern, can surely only be sought in the universe at large, so a common epistemological basis is a manifest desideratum.

A shape-based treatment breaks Rovelli's parallel between particles interacting with a piston in a box and gravitating particles in unbounded space. The behaviour of the particles is not what the piston analogy suggests. It is not entropic increase of disorder but creation of order, the essence of gravity. In unbounded space,  gravitating particles exhibit an arrow of time through dynamical necessity, not a special initial condition.

\section{$N$-Body Kinematics and Dynamics\label{b}}

The \nb has a phase space of unbounded measure while that of the particles and piston in a box is bounded.\footnote{Strictly, as will be detailed, a large set of \mb solutions have an unbounded measure. Some, but by no means all, negative-energy solutions explore only a bounded part of the \mb phase space.} This makes a big difference and is related to the fact that in the box model the piston's dof, which Rovelli uses to model the scale factor $a(t)$ in general-relativistic cosmological models, is a dof adjoined to the particle dofs, whereas in the \nb the scale dof is not an extra dof but a (mass-weighted) collective dof formed from all the $N(N-1)/2$ interparticle separations as the \e{root-mean-square length} \eln:\footnote{Up to a power of the constant total mass $M$, \el is identically equal to the centre-of-mass moment of inertia (half the trace of the inertia tensor).}
\be
\ell_\textrm{\scriptsize rms}={1\over M}\sqrt{m_im_jr_{ij}^2},~~r_{ij}= |{\bf r}_i-{\bf r}_j|,~~M=\sum_{i=1}^{i=N}m_i.\label{rms}
\ee
Moreover, since no external rod exists to measure the particle separations $r_{ij}$, they are determined only up to an overall scale. To pass to \e{scale-invariant} separations, one must divide the $r_{ij}$ by \eln, thereby reducing their number by 1.

This does not yet give a good way to measure the size of the universe using only intrinsic quantities---ones that do not rely on an implicit absolute scale. This problem can be overcome because besides \eln, to which the large $r_{ij}$ make the dominant contribution, there exists a further mass-weighted fundamental length in the \nbn, namely the \e{mean harmonic length} $\ell_\textrm{\scriptsize mhl}$:
\be
\ell_\textrm{\scriptsize mhl}^{-1}={1\over M^2}\sum_{i<j}{m_im_j\over r_{ij}}\propto -V_\textrm{\scriptsize New},~~V_\textrm{\scriptsize New}=-\sum_{i<j}{m_im_j\over r_{ij}}. \label{mhl}
\ee
It can be seen that $\ell_\textrm{\scriptsize mhl}$ is, with $G=1$, inversely proportional to minus the Newton potential $V_\textrm{\scriptsize New}$, which means that the small $r_{ij}$ make the dominant contribution to $\ell_\textrm{\scriptsize mhl}$. In measurement a short interval is always used to measure a long one. Since $\ell_\textrm{\scriptsize mhl}<\ell_\textrm{\scriptsize rms}$, one can say that $\ell_\textrm{\scriptsize mhl}$ measures $\ell_\textrm{\scriptsize rms}$ and that for any particle distribution
\be
{\ell_\textrm{\scriptsize rms}\over \ell_\textrm{\scriptsize mhl}}\label{scale}
\ee
is the intrinsic size of any \mb  model universe. Any change of scale determinable within such a universe reflects the behaviour of (\ref{scale}), which, like (\ref{rms}), is a collective dof and not some further dof as implicit in the treatment in \cite{rov}.

Because close approach or even coincidence of two or a few particles changes $\ell_\textrm{\scriptsize rms}$ little (if $N$ is moderately large) but $\ell_\textrm{\scriptsize mhl}$ a lot, (\ref{scale}) is a sensitive measure of clustering and variety. It is called the \e{shape complexity} $C_\textrm{\scriptsize shape}$ in \cite{bkm, jp}:
\be
C_\textrm{\scriptsize shape}={\ell_\textrm{\scriptsize rms}\over \ell_\textrm{\scriptsize mhl}}.\label{comp}
\ee
As will be shown below, secular growth of $C_\textrm{\scriptsize shape}$ gives rise to dynamical (not statistical) arrows of time in the \nbn. They are moreover arrows of increasing order, not disorder as the second law of thermodynamics implies.

Further evidence for the significance of (\ref{comp}) is that, as (minus) the normalised Newton potential or \e{shape potential} $V_\textrm{\scriptsize shape}$: 
\be
V_\textrm{\scriptsize shape}={1\over M^2}\ell_\textrm{\scriptsize rms}V_\textrm{\scriptsize New},\label{shp}
\ee 
it, rather than $V_\textrm{\scriptsize New}$, is what controls the \nb treated as a model universe. Indeed, a moment's reflection shows that it is (\ref{shp}) which counts, since an arbitrary overall scale of length (and mass) can only be nominal. It is striking that the purely mathematical measure of variety $C_\textrm{\scriptsize shape}$ (\ref{comp}) gives rise to and controls gravity. This is essentially a purely geometrical fact; mass ratios merely modify the gravitational effects already present for identical points in Euclidean space.

Matching the concepts above, it is appropriate, when treating \mb solutions as model histories of the universe, to replace the $3N$-dimensional \n configuration space ${\mathcal N}$ by the $3N-7$-dimensional \e{shape space} \s obtained from ${\mathcal N}$ by quotienting with respect to the similarity group {\sf Sim} of Euclidean translations, rotations, and dilatations:
\be
{\mathcal S}={{\mathcal N}\over{\sf Sim}}.\label{ss}
\ee

To reveal the creative core of gravity, we need a treatment of the \nb that involves only {\sf Sim}-invariant concepts, but to show how important is the scale-invariance enforced by the inclusion of dilatations in {\sf Sim} it is illuminating to prepare for that with concepts invariant under only the Euclidean group {\sf Euc}. The appropriate space in this case is the \e{relative configuration space} (RCS) ${\mathcal R}$,
\be
{\mathcal R}={{\mathcal N}\over{\sf Euc}},\label{rcs1}
\ee
which, unlike shape space \sn, has an unbounded measure (having been obtained without quotienting by dilatations). Whereas the complexity (or shape potential) stands out as a function on \sn, it is the Newton potential \isa that it is natural to consider as a function invariant under {\sf Euc} on the RCS. 

Besides these functions on the respective spaces, group theory also leads naturally to metrics on them through a procedure introduced in \cite{bb} to implement Mach's principle and called \e{best matching} in \cite{1994}. It is a universal group-theoretical procedure that defines change relationally. It can be applied in a particle or field ontology using groups that depend respectively on a finite number of parameters or arbitrary functions. In field theory it provides a common foundation for gauge theory and general relativity (see \cite{fl} for the general treatment). Here, I will consider only best matching for a finite number of particles in Euclidean space, which is all we need for the \nbn. The aim in that context is to define change of position without the notions of absolute space and time or, in their modern guise, an inertial frame of reference. 

Consider $N$ particles in Euclidean space. In the \n representation, infinitesimal changes $\textrm d{\bf r}_i$ from their initial positions ${\bf r}_i^0$ are the sum of intrinsic changes in the $r_{ij}$ and, quite independently, extrinsic changes generated by infinitesimal transformations of {\sf Sim} acting on the ${\bf r}_i^0$. The associated initial velocities are, moreover, undetermined up to an overall constant since they are determined through division by an unobservable increment $\textrm dt$ of absolute time. Considering first the geometrical extrinsic effects, they can be eliminated while retaining the advantage of a frame of reference that defines the initial positions ${\bf r}_i^0$. In one, consider first the \n kinetic metric
\be
\textrm ds^{\textrm{\scriptsize N}}=\sqrt{\sum_im_i\textrm d{\bf r}_i\cdot\textrm d{\bf r}_i},~~
\textrm d{\bf r}_i={\bf r}_i-{\bf r}_i^0,\label{my1}
\ee
corresponding to displacements from ${\bf r}_i^0$ to ${\bf r}_i$. It defines a line element in the Newtonian configuration space ${\mathcal N}$. To eliminate the extrinsic contribution to $\textrm ds^{\textrm{\scriptsize N}}$ inherent in the increments $\textrm d{\bf r}_i$ in (\ref{my1}), add to them arbitary infinitesimal increments generated by {\sf Euc} or {\sf Sim} transformations, obtaining trial increments.\footnote{They can, as in \cite{bb}, be expressed a gauge-invariant manner in terms of the generators of the Lie algebra of {\sf Euc} or {\sf Sim}, as here. The minimised quantity in (\ref{my}) is then seen to be the horizontal component of a gauge connection in a fibre bundle whose base is the RCS (or \sn). In fact, since the rotation group is non-Abelian, Einstein would have created the first and simplest non-Abelian gauge theory had he attempted to implement Mach's principle in particle dynamics by best matching. Conceptually, I prefer to present the notion of best matching without the gauge generators, which are auxiliary mathematical structures. Configurations, in the case of {\sf Euc}, or shapes in case of {\sf Sim}, are the reality; in the simplest three-body case, {\sf Euc} best matching can be illustrated by placing incongruent cardboard triangles relative to each other in the best-matched position.} Since (\ref{my1}) is positive definite and some intrinsic change is assumed, there must exist a `best-matching' set of them that mimimises (\ref{my1}) and in the case of {\sf Euc} gives a `Euclideanised' line element
\be
\textrm ds^{\textrm{\scriptsize{\sf Euc}}}_\textrm{\scriptsize bm}={{\textrm{min}}\over{\sf Euc}}\sqrt{\sum_im_i\textrm d{\bf x}_i\cdot\textrm d{\bf x}_i},~~
\textrm d{\bf x}_i={\bf x}_i^2-{\bf x}_i^1\label{my}
\ee
in which the centres of mass have been brought to concidence and there is no overall rotation. The line element (\ref{my}) defines geodesics in the RCS ${\mathcal R}$ (\ref{rcs1}). At this point, we recall Jacobi's action principle, which, for one fixed value $E$ of the energy, gives the timeless orbit of a dynamical system with potential $V$ in the \n configuration space ${\mathcal N}$ between configurations 1 and 2:
\be
A_\textrm{\scriptsize Jacobi}={1\over 2}\int_1^2\textrm d\lambda\sqrt{(E-V)T}, ~~T={1\over 2}\sum_im_i{\textrm d{\bf r}_i\over\textrm d\lambda}\cdot{\textrm d{\bf r}_i\over\textrm d\lambda}.\label{jac}
\ee
Here, $\lambda$ is a suitably continuous and monotonic but otherwise arbitrary parameter. As a result, the Jacobi action is invariant under reparametrization
\be
\lambda\rightarrow \lambda'(\lambda),~ \textrm d\lambda'/\textrm d\lambda>0.\label{rep}
\ee 
In normal physics applications, for example, in the Kepler problem, Jacobi's principle is used to find the orbit, after which energy conservation is used to find the speed in orbit. However, this requires an external clock to tell the time, which does not exist if the considered system is the entire universe. However, the equations of motion that follow from (\ref{jac}),
\be
{\textrm d\over\textrm d\lambda}\left({(E-V)\over T}\right){\textrm d{\bf r}_i\over\textrm d\lambda}=-\left({T\over(E-V)}\right){\partial V\over\partial{\bf r}_i},\label{eom1}
\ee
allow the introduction of a distinguished time parameter. This is because the value of $T$ can be freely changed by reparametrization in accordance with (\ref{rep}) and one can choose $\lambda$ to ensure that throughout the evolution
\be
E-V=T.\label{lab}
\ee
Then $(E-V)/T$ and its inverse both become unity, and the equations of motion (\ref{eom1}) take the form of Newton's 2nd law with the distinguished $\lambda$ now denoted by $t$:
$$
{\textrm d{\bf r}_i\over\textrm dt}=-{\partial V\over\partial{\bf x}_i}.\label{eom2}
$$
Of course, rewritten in the form $E=T+V$ the implicit definition (\ref{lab}) of the distinguished $\lambda$ looks like the statement of energy conservation, but it is better interpreted in the first place as the definition of the independent variable that casts the equations of motion into their simplest form. In astronomical practice, it is \e{ephemeris time} \cite{bb, not}. We will see later that clocks form that are good in the sense that they all tell the same time as each other and the time that (\ref{lab}) defines.

This is a good point for an intermediate summary. From the point of view of defining geodesics on the RCS by means of best matching, the simplest possibility is to set $E$ equal to zero in (\ref{jac}) and thereby, in a very natural way, eliminate the ambiguity noted above that is associated with the pairing of an arbitrary $\textrm dt$ with the geometrical increments $\textrm d{\bf x}_i$. Then choice of the orbit parameter $\lambda$
by means of the simplifying condition (\ref{lab}) leads to the emergence of Newton's time from a timeless theory.  

This is not all that the theory achieves. Best matching permits horizontal and vertical stacking \cite{bb}, in which each successive configuration (any one of which may be taken to be realised at a nominal origin of time) is supposed `laid horizontally on top' of its predecessor in the best-matched position with `vertical spacing' between successive configurations fixed by the ephemeris time. In the framework that is thereby constructed, Newton's laws are satisfied exactly as employed today in inertial frames of reference.\footnote{Despite its foundational status, the origin of inertial frames is seldom if ever addressed in textbooks. At best, it is simply said they are frames in which Newton's laws take their usual form.} Besides the bonus of the relational derivation of inertial frames of reference, three Machian predictions are made: the total energy, momentum, and angular momentum of the \mb universe all vanish. All three come from the relational requirement that nothing apart from separations between particles (and mass ratios) should appear in the theory. 

We now come to the most interesting point, which is what happens if we attempt to extend the success achieved in the RCS to shape space \s and make the theory scale-invariant. It is worth emphasising here the attraction of geodesic theories. They are \e{maximally predictive} in the sense that just two points in any metric space determine a geodesic and with it a solution to the equations of the theory. The question then arises of what kind of geodesic theories can be defined in \sn. The critical step to that is, prior to best matching, to modify the \n kinetic metric (\ref{my1}), which neither on its own nor multiplied by \isa on its own or $E-V_\textrm{\scriptsize New}$ as in (\ref{jac}), is scale-invariant.\footnote{In the timeless Jacobi action (\ref{jac}), $E$ and \isa have dimensions [mass]$^2$[length]$^{-1}$.}

We obtain the simplest best-matched and scale-invariant action if instead of acting on the square root in (\ref{my}) we act on
\be
\sqrt{{\sum_im_i\textrm d{\bf x}_i\cdot\textrm d{\bf x}_i}\over\ell_\textrm{\scriptsize rms}^2}.\label{km}
\ee
Then the resulting minimum is scale-invariant and, since it does not involve forces, defines in \s what may be called \e{the natural metric}:
\be
\textrm ds_\textrm{\scriptsize bm}^{{\textrm{\scriptsize{\sf Sim}}}}={\textrm{min}\over{\sf Sim}}\sqrt{{\sum_im_i\textrm d{\bf x}_i\cdot\textrm d{\bf x}_i}\over\ell_\textrm{\scriptsize rms}^2}.\label{nm}
\ee
In the geodesics that result from this theory, which can be said to define `inertial motion' of the \mb universe in its shape space, we now have vanishing of not only the energy, momentum, and angular momentum but also a new quantity $D$,
\be
D=\sum_i{\bf p}_i\cdot{\bf x}_i,\label{dm}
\ee 
which in standard \mb theory is equal to half the time derivative of the centre-of-mass moment of inertia, which, like \eln, measures the size of the \mb universe. The size of such a universe is therfore constant. It does not expand. The same problem arises if prior to best matching we replace (\ref{km}) by
\be
\sqrt{{\sum_im_i\textrm d{\bf x}_i\cdot\textrm d{\bf x}_i\over V_\textrm{\scriptsize New}^{-1}\ell_\textrm{\scriptsize rms}}},\label{kmbis}
\ee
which, as shown in \cite{2003}, introduces \n gravitational forces but together with further forces that enforce constancy of the size of the universe, again because $D$ vanishes. The quantity $D$ (\ref{dm}) has the same dimensions as angular momentum and as a measure of overall expansion (as opposed to overall rotation) may, as in \cite{2003}, be called the \e{dilational momentum}. It appears often in \mb calculations but without a name. This is probably because it is only conserved with a potential that is homogeneous of degree $-2$, unlike \isan, which is homogeneous of degree $-1$. Best matching based on (\ref{km}) or (\ref{kmbis}) enforces $D=0$ and allows conservation of this vanishing value.

It is clear that the above attempt at a scale-invariant relational theory must be rejected; a universe which does not expand fails to match observations. Besides this failure, there is a further flaw. The quotienting with respect to dilatations makes \s compact and this means that all geodesic theories in \s are subject to \po recurrence and cannot provide a satisfactory theory of time's arrow.\footnote{I am grateful to Pooya Farokhi for alerting me to this fact.} However, this second problem with the above attempt at scale-invariance points the way to a quite different realisation of the aspiration. A hint in that direction was first obtained in \cite{bkm}, in which it was shown that a large set of \mb solutions, indeed all in which the energy is nonnegative, exhibit dynamical arrows of time defined by secular increase of the complexity (\ref{comp}). This result, and its bearing on \cite{rov}, will now be presented.

\section{Generic Behaviour of $N$-Body Solutions\label{c}}

Any \n \mb solution obtained in ${\mathcal N}$ can be projected to shape space \s as an unparametrised undirected curve of successive shapes. Without loss of any physical information, this eliminates all redundant structure that might be mistaken for reality, for example, the direction implied by Newton's notion of absolute time and an absolute scale determined by a ruler outside the universe.\footnote{Whereas the conflict between the second law and time-reversal symmetry has generated a vast literature, little has been addressed to  scale (in three dimensions, not to the as yet fruitless attempt for conformal symmetry in spacetime) except in the relatively recent shape-dynamic approach, of which \cite{fl} provides an extensive introduction.} Moreover, the \n solutions can always be recovered from the curves in \s \cite{knv}. In \cite{grub} all \mb solutions are classified according to their different possible behaviours. 

As shown in \cite{bkm}, the most significant fact in the present context is that all \mb solutions with nonnegative (centre-of-mass) energy $E$ have arrows of time unrelated to entropic statistics and macrostates. In every microhistory, secular growth of $C_\textrm{\scriptsize shape}$ (\ref{comp}) defines a direction. If \el (\ref{rms}) does not vanish, bidirectional arrows exist either side of a `Janus point' of a unique minimum of \eln, from which it increases monotonically to infinity in both time directions.\footnote{A subset of the negative-energy solutions exhibit virialised thermodynamic behaviour without an arrow of time while others have `Janus regions' and bidirectional arrows \cite{grub}.} There are also zero-measure solutions that exist on only one side of a point at which $\ell_\textrm{\scriptsize rms}$ has the value zero, from which it increases monotonically to infinity. They are `big-bang' solutions and most closely correspond to the Rovelli--Wallace \cite{rov, wall} project of modelling time's arrow in the universe by means of the \nbn. However, Janus-point solutions are ubiquitous in the \mb phase space. Their existence and properties are sufficient for my purposes and highlight the difference between confined thermodynamic systems and unconfined \mb systems. Some of the fascinating properties of the `big-bang' solutions and the behaviour of $C_\textrm{\scriptsize shape}$ ({\ref{comp}) in them are discussed in \cite{jp, fl-p, may}.

\begin{figure}
\begin{center}
\includegraphics[width=0.6\textwidth]{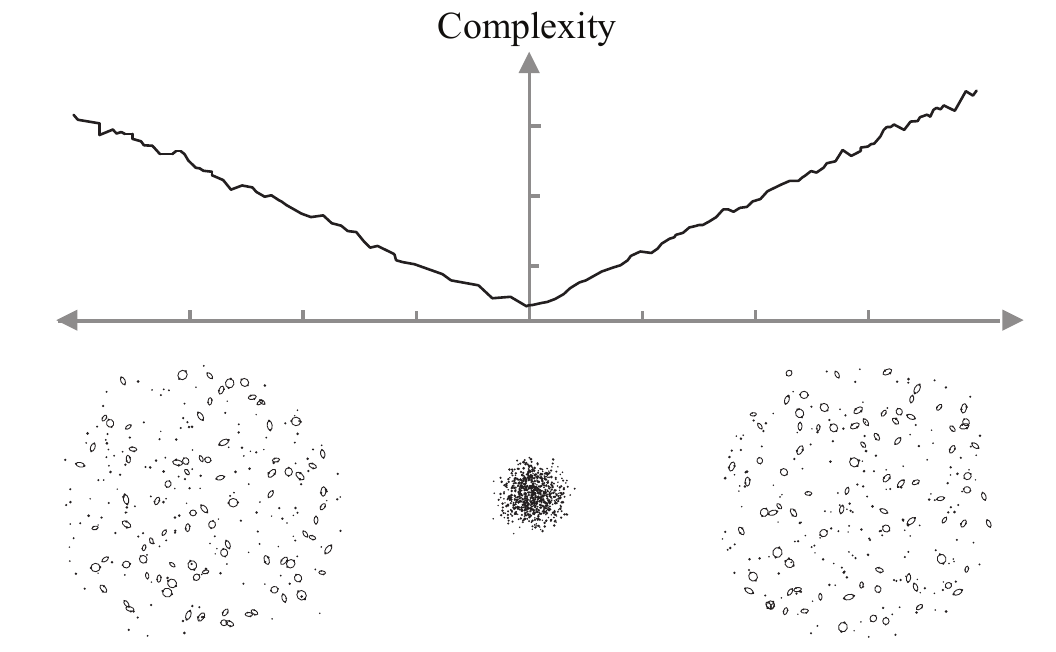}
\caption{\small Growth of \c either side of the Janus point in a numerical solution for 1000 equal-mass particles \cite{bkm} and `artist's impression' of the system's appearance: very uniform in the region of the Janus point and clustered either side of it.}\normalsize\label{fig1}
\end{center}
\end{figure}

Figure 1 is a typical \mb solution with $E\ge 0$ for which $\ell_\textrm{\scriptsize rms}$ does not vanish.\footnote{The solution has $E=0$ and vanishing angular momentum. As I will explain, these conditions, which model closed-space cosmological models in general relativity, reveal the `creative core' of \mb theory without changing the qualitative behaviour that all solutions with $E\ge 0$ share.} The complexity grows between linearly rising bounds either side of a Janus point of minimum $\ell_\textrm{\scriptsize rms}$. The state of the system in the vicinity of the Janus point corresponds at least qualitatively to high entropy: the particles have a more or less uniform distribution and random velocities. In the `big-bang' solutions discussed in \cite{jp}, chaps. 16--18, and \cite{may} the qualitative behaviour is broadly the same as half of the solution from the Janus point. From Fig.~1 we conclude: 

\begin{enumerate}

\item Arrows of time exist in a single solution. This refutes the comment in \cite{rov} that ``[It] is essential to recall that irreversibility is a macroscopic notion. It is a property of a certain coarse graining. It is not a property of a microscopic dynamical evolution.'' This is true for confined systems but not as shown above for the unconfined \nbn. The growth of complexity in Fig.~1 either side of the Janus point is irreversible and a microscopic dynamical evolution. 

\item Neither a `past hypothesis' \cite{a} nor low entropy in the past \cite{rov} is needed for a Janus point. Its existence follows directly from Newton's second law \cite{bkm} and the negative-definiteness and homogeneity of degree $-1$  of $V_\textrm{\scriptsize New}$.\footnote{This homogeneity and the associated dynamical similarity is the basis in \cite{sl} of much the same content as presented here. Potentials besides $V_\textrm{\scriptsize New}$ for which Janus points exist are listed in \cite{ad}.} 

\item On the basis of intrinsic facts---the $r_{ij}/\ell_\textrm{\scriptsize rms}$ and their evolution---the system, with most uniform spatial distribution of its particles and more or less random velocities of them, is in its highest, not lowest, entropic state at the Janus point---the past for internal observers.\footnote{Wallace, with whom Rovelli agrees, uses known facts in \mb theory to argue, like many theoreticians (above all Penrose \cite{pene, penr}), that uniform states in gravity have low entropy. However, Wallace's most persuasive examples are globular clusters, which, following formation within the universe through processes that cannot be described in \mb theory, have negative energy and therefore do not meet the condition $E\ge 0$. Moreover, they slowly `decay' through the evaporation of individual stars, during the process of which the complexity of the complete system formed by the stars still in the cluster together with the `escapees' behaves as it does in either half of Fig.~1. Like the \nbn, globular clusters (once formed) have an unbounded phase space.} 

\item For all solutions with $E\ge 0$, the behaviour of $C_\textrm{\scriptsize shape}$ away from the Janus point does not correspond to growth of disorder but order. This is seen most clearly in the solutions with $E=0$ and angular momentum ${\bf L}=0$. Being conserved, these quantities remain zero for the complete system, but away from the Janus point the more or less isolated subsystems depicted in Fig.~1 form with all possible approximately conserved $E,~ {\bf L},$ and momenta ${\bf P}$, their respective sums being zero. This creation of order is especially pronounced in `Kepler pairs': two particles with evermore perfectly elliptical orbits around their centre of mass. Through their periods and major-axis lengths and directions they spontaneously synchronise with one another, each as an evermore accurate clock, rod, and compass all in one. Once formed, nearly all exist forever. 

\item As noted in \cite{grub}, the subsystems, being more or less well isolated while they exist, exhibit thermodynamic behaviour. In particular, in being virialised, they approximate \po recurrence. Moreover, they all have mutually aligned birth-virialisation-death arrows of time that point in the same direction as the complexity arrow on their side of the Janus point. The complexity arrow is therefore a dynamical master arrow, while the arrows of the subsystems are subordinate, thermodynamic, and emergent within each microhistory.\footnote{In 1962 Gold \cite{go} argued that expansion of the universe, by ensuring the night sky is dark and has a corresponding low temperature, is the master arrow of time behind all the others, including that of entropy. The problem of whether and how entropy might decrease should the universe start to recontract to a big crunch was an obvious problem. In the \mb solutions discussed here it does not arise because their expansion, measured both in \n absolute terms and intrinsically by the complexity, continues forever. Note also that the increase of the universe's size as measured intrinsically by (\ref{scale}) and the growth of \c (\ref{comp}) are one and the same thing. This basic-level unification is not present in Gold's proposal.\label{gol}}

\end{enumerate}

\section{The Creative Core of $N$-Body Dynamics\label{d}}

There are two main reasons for the differences from long held beliefs about the origin and nature of time's arrow itemised above: the unbounded phase space of the \nb and the role of $V_\textrm{\scriptsize New}$ in its scale-invariant form \npn. They combine to reveal something that does not seem to have been hitherto recognised and is the subject of this section: a creative core of \mb solutions. It is hidden not only in the `boxed gravitational catastrophe' \cite{rus}, used in \cite{wall} to model globular clusters, but also in all systems, whether gravitating or not, confined by a physical container or nondynamical potential. Even in systems free of these artificial constraints the absolute notions of time, place, orientation, and scale that Newton added to the bare notions of Euclidean geometry and effectively remain in inertial frames of reference hide the core to a considerable extent. Another factor is the development of dynamics, whose origins in astronomical prediction led to concentration on finite-time evolution from initial conditions. Such work cannot reveal the creative core, which requires study of generic properties of complete solutions.

The reader will naturally ask what I mean by the creative core. One possibility could be that it consists of the set of \mb solutions that, in a well-defined sense, are maximally predictive and, to satisfy the `Heisenberg aspiration', are defined in intrinsic terms and thus in shape space \sn. As already noted in Sec.~\ref{b}, such a theory will be geodesic since then a point and direction in \s determine a solution; no theory can do better than that. However, the conclusion reached at the end of Sec.~\ref{b} was that any such geodesic theory suffers from flaws, one fatal and the other undesirable. Moreover, the root of both problems is scale-invarance, just what we wanted! The question now arises of whether some other scale-free approach, not necessarily based on the criterion of maximal predictive strength, can yield what we seek. The two most important things are what the maximal-predictability criterion ruled out: an expanding universe and an arrow of time. From this point of view, the solutions considered in Sec.~\ref{c} are promising since in them the behaviour of the complexity meets both desiderata: it defines not only a dynamical arrow of time but also growth of the intrinsic observable size, doing so moreover in a manner that unifies fundamental concepts since, as noted in Sec.~\ref{b} and footnote~\ref{gol}, the intrinsic scale (\ref{scale}) and the complexity (\ref{comp}) are one and the same function.

Let us now consider which among all \mb solutions come closest to the predictive ideal while still meeting both desiderata. The answer reveals the effect of Newton's absolutes. Generic \mb solutions fall short of the ideal because their Cauchy data contain five dimensionless global quantities that cannot be encoded in a shape and direction in \sn. First, by the velocity decomposition theorem \cite{saari}, the kinetic energy $T$ at any instant decomposes into the components in change of shape, rotation, and expansion (dilatation): $T_\textrm{\scriptsize s}$, $T_\textrm{\scriptsize r}$, and $T_\textrm{\scriptsize d}$. The ratios $T_\textrm{\scriptsize r}/T_\textrm{\scriptsize s}$ and $T_\textrm{\scriptsize d}/T_\textrm{\scriptsize s}$ are two missing data. Two more fix the direction of ${\bf L}$ relative to the instantaneous shape. The final datum is $T/V_\textrm{\scriptsize New}$.

Now except for $T_\textrm{\scriptsize d}/T_\textrm{\scriptsize s}$, which appears because the \n scale variable \el is not constant, the other four missing data play no role if $E$ and $L$ are zero as they are for Machian solutions that, as shown in \cite{bb} and explained above, are free of any effects of absolute position and orientation. In the traditional conceptual framework, $T_\textrm{\scriptsize d}/T_\textrm{\scriptsize s}$ therefore appears to be the `last vestige of Newton's absolutes'. It is not in scale itself but in a ratio of scales at different times. Note that this statement is expressed in \n terms. It still employs the notion of an external ruler used to measure the \n size \el of the universe at different times. 

Things look different if considered entirely in shape space \sn. In such a context, only concepts used in the definition of \s can be employed. They are ratios of distances between mass points and ratios of their masses, though, as noted earlier, the latter do not change the essential features of the \nb but only modify their expression. From the basic shape-space `ingredients' one can form functions on \sn. As our primary concern is with global properties, these functions should involve all the particles and be mass-weighted. Since $3N-7$ dofs define any nondegenerate $N$-particle shape, that is maximal number of appropriate collective functions. The simplest and most obvious one to consider is the complexity. This is not the place to consider other such functions, which could, for example, take the form 
$$
C_\textrm{\scriptsize shape}={\ell_\textrm{\scriptsize rms}^n\over \ell_\textrm{\scriptsize mhl}^n}, ~~n=2\dots~3N-7,
$$
and, together with \cn, serve as configurational state functions. 

Besides such state functions, curves can be defined in \sn. In particular they can represent \n \mb solutions projected to \s as described earlier. As the above arguments showed, none of the resulting curves will be geodesic, but it is interesting to consider the manner in which they are nongeodesic in the light of what Leibniz called his two great principles: the \e{identity of indiscernibles}, according to which two things that are supposed to be ontologically distinct but differ in none of their attributes must be recognised to be the same thing, and the \e{principle of sufficient reason}, according to which any (observable) effect must have a definite cause. The identity of indiscernibles is the justification for passing from the \n configuration space ${\mathcal N}$ to \sn; indeed, Leibniz implicitly defined at least the RCS (\ref{rcs1}) in his correspondence with Clarke \cite{lc} when, clearly anticipating Mach, he said explicitly that the only motion that the universe can meaningfully be said to have is ``as its parts change their situation among themselves''. A sufficient reason, for its part, was used above to argue that the constant $E$ should not appear in the formulation of Jacobi's principle since nothing in the bare concept of a path in the RCS (\ref{rcs1}) could explain why an extraneous $E$ with any value should appear in (\ref{jac}).

The principle of sufficient reason can also be used to reject \mb solutions with nonvanishing angular momentum ${\bf L}$. Although they can be described in \s by introducing three additional degrees of freedom, no reason at all can be found in \s why the behaviour to which ${\bf L\ne 0}$ leads should arise (though a reason appears for subsystems since they form with some relation to the complete \mb universe). However, the situation is different with regard to the `last vestige of Newton's absolutes when we consider how, despite the failure of maximal predictability, the curves of $E={\bf L}=0$ solutions behave in \sn. A sufficient reason for their behaviour can be identified and even depicted in the case of the \tb through the two images in Fig.~\ref{fig2} (created by Flavio Mercati).

\begin{figure}
\begin{center}
~~\includegraphics[width=0.285\textwidth]{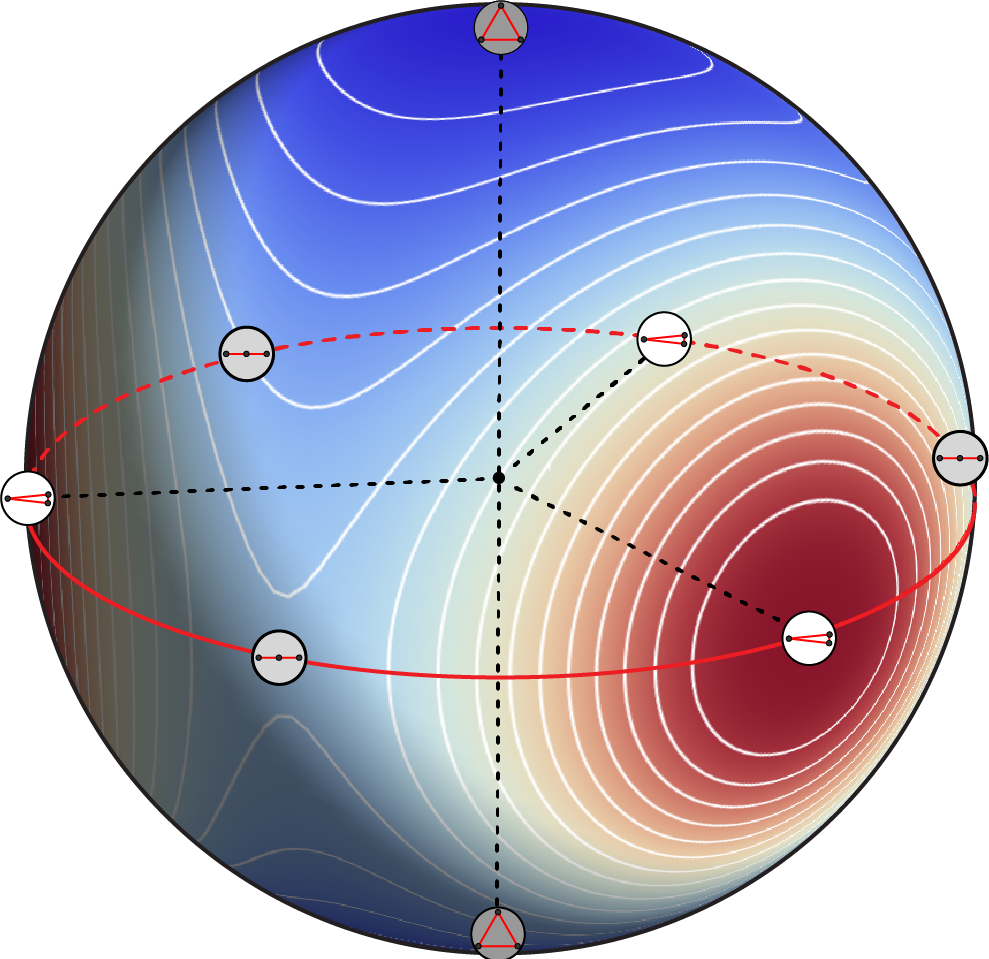}~~~~~~~~~~~~~\vspace{-.1in}\includegraphics[width=0.28\textwidth]{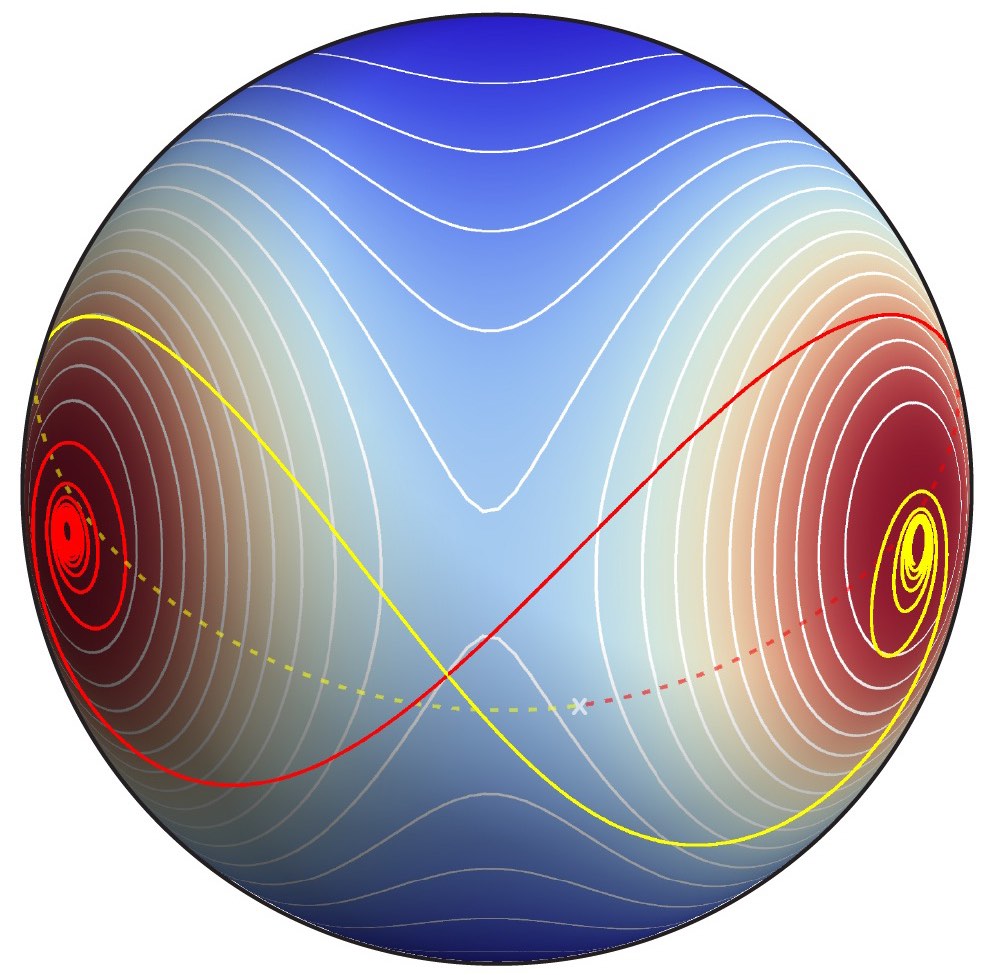}
\caption {\small The shape sphere (left) and a three-body solution with $E={\bf L}=0$ (right).}\normalsize\label{fig2}
\end{center}
\end{figure} 

The one on the left exploits the fact that in the \tb the three particles are, at any instant, at the vertices of a triangle. To implement the `Heisenberg aspiration', we express everything in terms of the successive triangle shapes, defining each by two internal angles. Suitably mass weighted, they can be represented by points on the surface of a \e{shape sphere}, as on the left in Fig~\ref{fig2}. The natural metric (\ref{nm}) determines the distance on the surface of the sphere, which, with contours, is colour-coded by the value of \cn. Chiral triangles are at equal longitudes and opposite latitudes. Collinear triangles are on the equator, on which there are three saddles of \c named after Euler, who first discovered their significance, and three two-body coincidences, at which \c is infinite. In the equal-mass case shown in Fig.~\ref{fig2} the absolute minimum of \c (always at the equilateral triangle whatever the masses) is at the poles but is displaced from them if the masses are unequal.

We now consider the possible \n evolutions within this kinematic arena. Apart from the zero-measure set of `big-bang' solutions discussed in \cite{jp}, chapters 16--18, \cite{fl-p}, all three-body solutions with vanishing energy and angular momentum asymptote on either side of a Janus point to hyperbolic--elliptic escape, in which two particles orbit their common centre of mass in Keplerian ellipses, while a `singleton' flies off in the opposite direction with energy and momentum that balance the Kepler pair's.\footnote{All the noncollinear zero-measure solutions start at the equilateral triangle and also asymptote to a Kepler pair and singleton. All three-body solutions with ${\bf L}=0$ are planar.} In shape-dynamic terms, the behaviour, illustrated on the right in Fig.~\ref{fig2}, has a simple interpretation. The system has an innate tendency to move along a geodesic of the natural metric (\ref{nm}), in this simplest three-body case a great circle on the shape sphere. However, the system is also under the influence of the shape potential $V_\textrm{\scriptsize shape}$, which up to its sign is equal to the complexity $C_\textrm{\scriptsize shape}$. As a result, as the system's representative point recedes along either branch from the Janus point (shown by the white cross at the back of shape sphere) it `strives', with fluctuations, to climb ever higher up one of the complexity peaks. \e{Striving to greater complexity} is the eminently Leibnizian sufficient reason for the deviation from the simplest geodesic motion in \sn. It is what I call the creative core of \n gravity. It occurs whatever the value of $N$.\footnote{As pointed out in \cite{jp}, chapter 12, Newton did not create the theory of a clockwork universe with cogs that turn mindlessly; his is actually a theory of creation. Item 4 of conclusions in Sec.~\ref{c} shows how, if the direction of increasing complexity is identifed as the direction of time, matching what we experience in the universe, then the \nb creates out of chaos clocks that then, all synchronised with one another, `tell time' with ever better accuracy.} If mathematicians with no knowledge of gravity were presented with the concept of shape space and asked to develop a nontrivial theory of curves within that arena they could hardly fail to come up with curves like the one on the right in Fig.~2.

That figure would differ significantly but not entirely lose its nature were the energy positive and/or the angular momentum nonzero. There would still be `striving' to higher \c either side of a Janus point, whose existence only requires the energy to be nonnegative. However, by allowing the four extra terms to play a role in the dynamics, absolute time and space weaken the striving. The effect of a nonvanishing energy $E$ suffices to make the point. As noted in Sec.~\ref{c}, some solutions with $E<0$ never leave a bounded region of the \mb phase space; for them, but by no means all solutions with $E<0$, striving to higher complexity never happens. If $E>0$, there is always striving but some solutions can asymptote in one or both branches of the solution to hyperbolic escape. This is when all the particles escape to infinity and the evolution in \s freezes at some value of \cn; it is only in the extended \n representation that, as a gauge effect, it continues. Even if the shape does not freeze, the system's representative point is forced to deviate from the most direct route to higher \cn. Angular momentum has the same effect.

Whereas the `absolutes' weaken the growth of \cn, confinement stops it, brutally one can say. Even if nothing in \s gives a sufficient reason for their existence, the `absolutes' are at least manifested through dynamical dofs that interact in a deterministic manner with the shape dofs. Confinement, in contrast, changes the system's motion, abruptly and unnaturally, whenever particles encounter the wall of a box or a nondynamical potential barrier. The combined effect is to prevent growth of \el above a certain upper limit, around which it hovers forever, with only rare dips to lower values, very exceptionally to zero. The behaviour of \cn, while bounded below and subject to brief jumps of arbitrary height induced by close encounters of two or a few particles, is much the same. It's no surprise that both these behaviours are much like those of entropy in confined systems.

Now suppose how `mathematicians from another universe', able to `see' only as Heisenberg advocated for quantum effects, might formulate the law that governs the solutions that belong to the creative core. Unable to see the \n kinematic framework but only \s and, in it, evolution curves with a never ending tendency to bend away from the natural geodesics to higher complexity, it would never occur to them to invoke ratios of scale, let alone scale. Like the advocates of pure shape dynamics \cite{knv}, they would, to describe the bending away from the natural geodesics in \sn, simply add to the $6N-15$ shape and shape-direction dofs one more: the variable $\kappa$ employed as measure of the bending in \cite{knv}. Because the single extra bending variable reacts solely and directly to the scale-invariant shape potential \npn, it would by no means appear to be out of place in \sn. Interpreted in the proper context, the `last-vestige puzzle' disappears. Creation is seen for what it is. 

However, the above discussion still lacks something from the point of view of sufficient reason. All the solutions with a Janus point have in the \n representation a finite value of \eln.  As shown in \cite{knv}, the location in shape space \s of the Janus point is where the bending variable $\kappa=2C_\textrm{\scriptsize shape}$. This means it can be anywhere in \sn; no sufficient reason can be found for the location of the Janus point in any solution that has one. If our model \n universes are to satisfy Leibniz's principle, we need to consider the `big-bang' solutions. They are like `half' of a Janus-point solution but begin at special points in \s called central configurations, at which the value of \c is extremal (either at a saddle or minimum). For such solutions, the problem with a sufficient reason is to a large extent eliminated since all solutions begin at very special points in \sn. Indeed, there is in general a unique central configuration, which may be called \e{Alpha}, at which \c has its smallest possible value (being positive definite and defined on a compact space, \c must have an absolute minimum). Thus, for all $N$ for which there is a unique Alpha, just one initial shape is singled out by Leibniz's principle. Consideration of the radical differences and possibilities that come into view when the dynamics of the universe is examined in the light of this fact go beyond the scope of this paper but are discussed in the online talk \cite{may}.

\section{Concluding Comments}

In this final section, I want to discuss the status of the second law of thermodynamics in cosmology and begin with Eddington's famous statement

\begin{quote}\small

The law that entropy always increases holds, I think, the supreme position among the laws of Nature \dots [If] your theory is found to be against the Second Law of Thermodynamics I can give you no hope; there is nothing for it but to collapse in deepest humiliation.

\end{quote}\normalsize

This kind of conviction seems to be behind virtually all discussions of the second law in the context of the universe. It explains the felt need for either some addition to the known laws of nature in the form of a past hypothesis, which imposes by brute force a condition of low entropy in the past, or the proposal, advocated especially by Penrose and followed widely, that gravitational entropy, unlike matter entropy, increases and not decreases with clustering. It should be clear from the previous discussion that I question this conclusion. It ignores an elephant in the room: are we considering systems that have phase spaces of bounded or unbounded measure? Eddington's comment may be compared with Einstein's on thermodynamics \cite{ein}: ``It is the only physical theory of universal content which I am convinced that, within the framework of applicability of its basic concepts, will never be overthrown.'' 

Einstein did not spell out the framework of applicability---it is hard to find that done anywhere---but I think the arguments given in this paper show that confinement, either physical or conceptual, is its \e{sine qua non}. In view of its huge departure from thermal equilibrium and observed accelerated expansion, the universe itself does not appear to be confined. As regards confined subsystems, I already noted that in the \nb more or less isolated subsystems can and do form. If they contain as many particles as the Galaxy, they will, like black holes, survive for a very long time and exist within a relatively small region of the universe. They will therefore satisfy Einstein's condition for meaningful (though never more than pragmatic) application of thermodynamic concepts. 

My tentative conclusion is therefore that, despite what Eddington said, the universe is not subject to the second law but that subsystems which form within it can, with suitable care, be described thermodynamically with some degree of accuracy. The conclusion is tentative since it is based on the behaviour of the \nbn, which may not extend to the universe, though what we observe in it does suggest that could well be the case. If it is, then to the extent to which emergent subsystems can be ascribed an entropy the arrow of time associated with it will always be aligned with the master arrow that a suitably generalised notion of complexity defines.\footnote{Possible generalisations of \c to a universe described by general relativity and the known matter fields are discussed in \cite{jp}, chapter 18.} The growth of entropy is subordinate to the complexity arrow.

A final question is this: what scientific predictions might be deduced from exploration of the shape-dynamic perspective? Might it be more fruitful than the science of confinement: thermodynamics and statistical mechanics? The rich harvest they gave in the 19th century, culminating in its very last months with the discovery of the quantum of action, can be seen as brilliant vindication of the view attributed (possibly not quite accurately \cite{pesic}) to the empiricist Francis Bacon in the early 17th century that it is necessary to `torture' nature to extract her secrets. Since thermodynamics and its statistical-mechanical interpretation both depend on confinement (`prison') and application of pressure, heat, and magnetic or electric fields (`screws') it can, metaphorically, be said that the secrets of nature discovered in the 19th century were indeed obtained by torture. However, besides welcome secrets, nature also indicated that prison leads to destruction of order and heat death. Perhaps nature must, as in the \nbn, be released from prison if she is to reveal her greatest secret: the capacity to create order with ever greater precision. That, at least, is what gravity's creative core suggests. Further hints in that direction are discussed in \cite{may}.

\vspace{.2in}

\e{Acknowledgements.} I'm particularly grateful to my long-term collaborators Tim Koslowski and Flavio Mercati, with whom the key ideas presented here have largely been developed. Flavio has also kindly made many other figures for me besides Fig.~2; Jerome Barkley created the complexity plot in Fig.~1 on the basis of calculations he made. Comments by Pooya Farokhi and Pedro Naranjo on earlier drafts of this paper have improved it usefully in several respects. I have also appreciated discussions with Anish Bhattachary and Kartik Tiwari.

\end{document}